\documentclass[journal,hidelinks]{IEEEtran}

\usepackage{graphicx,url}
\usepackage[english]{babel}
\addto\captionsenglish{}
\usepackage[utf8]{inputenc}
\usepackage{amsmath}
\usepackage{amssymb}
\usepackage{lipsum}
\usepackage{mathtools}
\usepackage{cuted}
\usepackage{cite}
\usepackage{graphicx}
\usepackage{balance}
\usepackage{tikz}
\usepackage{pgfplots}
\usepackage{tikzscale} 
\pgfplotsset{compat=newest} 
\usepackage{scalefnt}
\usepackage{orcidlink}
\usepackage{dblfloatfix}
\usepackage[acronym,shortcuts]{glossaries}
\usepackage{algorithm, algpseudocode}
\usepackage{algcompatible}
\usepackage{bm}
\usepackage{caption}
\usepackage{subcaption}
\usepackage{booktabs}
\usepackage{float}
\usepackage{dblfloatfix}

\newacronym{RMSE}{RMSE}{root mean square error}
\newacronym{MMSE}{MMSE}{minimum mean square error}
\newacronym{MF}{MF}{matched filter}
\newacronym{RPE}{RPE}{radar parameter estimation}
\newacronym{OTFS}{OTFS}{Orthogonal Time Frequency Space}
\newacronym{AFDM}{AFDM}{Affine Frequency Division Multiplexing}
\newacronym{MIMO}{MIMO}{multiple-input multiple-output}
\newacronym{SISO}{SISO}{single-input single-output}
\newacronym{ISAC}{ISAC}{Integrated Sensing and Communications}
\newacronym{3D}{3D}{three-dimensional}
\newacronym{2D}{2D}{two-dimensional}
\newacronym{1D}{1D}{one-dimensional}
\newacronym{RX}{RX}{receiver}
\newacronym{TX}{TX}{transmitter}
\newacronym{BF}{BF}{beamforming}
\newacronym{mmWave}{mmWave}{millimeter-wave}
\newacronym{SotA}{SotA}{state-of-the-art}
\newacronym{ULA}{ULA}{uniform linear array}
\newacronym{QAM}{QAM}{quadrature amplitude modulation}
\newacronym{ISFFT}{ISFFT}{inverse symplectic finite Fourier transform}
\newacronym{SFFT}{SFFT}{symplectic finite Fourier transform}
\newacronym{AWGN}{AWGN}{additive white Gaussian noise}
\newacronym{OFDM}{OFDM}{orthogonal frequency division multiplexing}
\newacronym{OCDM}{OCDM}{orthogonal chirp division multiplexing}
\newacronym{BS}{BS}{base station}
\newacronym{UE}{UE}{user equipment}
\newacronym{DFT}{DFT}{discrete Fourier transform}
\newacronym{IDFT}{IDFT}{inverse discrete Fourier transform}
\newacronym{IFFT}{IFFT}{inverse fast Fourier transform}
\newacronym{TD}{TD}{time-domain}
\newacronym{wlg}{wlg}{without loss of generality}
\newacronym{CP}{CP}{cyclic prefix}
\newacronym{DAFT}{DAFT}{discrete affine Fourier transform}
\newacronym{DAF}{DAF}{discrete affine Fourier}
\newacronym{IDAFT}{IDAFT}{inverse discrete affine Fourier transform}
\newacronym{CPP}{CPP}{\textit{chirp-periodic} prefix}
\newacronym{IDZT}{IDZT}{inverse discrete Zak transform}
\newacronym{DZT}{DZT}{discrete Zak transform}
\newacronym{ICI}{ICI}{inter-carrier interference}
\newacronym{BER}{BER}{bit error rate}
\newacronym{DoF}{DoF}{degrees-of-freedom}
\newacronym{FD}{FD}{full-duplex}
\newacronym{SIMO}{SIMO}{single-input multiple-output}
\newacronym{MISO}{MISO}{multiple-input single-output}
\newacronym{AoD}{AoD}{angle-of-departure}
\newacronym{AoA}{AoA}{angle-of-arrival}
\newacronym{RF}{RF}{radio frequency}
\newacronym{SIM}{SIM}{stacked intelligent metasurfaces}
\newacronym{FPGA}{FPGA}{field programmable gate array}
\newacronym{UPA}{UPA}{uniform planar array}
\newacronym{CC}{CC}{communication-centric}
\newacronym{I/O}{I/O}{input-output}
\newacronym{iid}{i.i.d.}{independent and identically distributed}
\newacronym{IoT}{IoT}{internet of things}
\newacronym{V2V}{V2V}{vehicle-to-vehicle}
\newacronym{V2X}{V2X}{vehicle-to-everything}
\newacronym{UAV}{UAV}{unmanned aerial vehicle}
\newacronym{NTN}{NTN}{non-terrestrial network}
\newacronym{LEO}{LEO}{low-earth orbit}
\newacronym{THz}{THz}{terahertz}
\newacronym{EM}{EM}{expectation maximization}
\newacronym{RIS}{RIS}{reconfigurable intelligent surface}
\newacronym{DoA}{DoA}{direction-of-arrival}
\newacronym{DD}{DD}{doubly-dispersive}
\newacronym{ODDM}{ODDM}{orthogonal delay-Doppler division multiplexing}
\newacronym{LoS}{LoS}{line-of-sight}
\newacronym{NLoS}{NLoS}{non-line-of-sight}
\newacronym{6G}{6G}{sixth generation}
\newacronym{MPDD}{MPDD}{metasurfaces-parametrized DD}
\newacronym{GaBP}{GaBP}{Gaussian belief propagation}
\newacronym{MSE}{MSE}{mean-squared-error}
\newacronym{sIC}{soft IC}{soft interference cancellation}
\newacronym{soft RG}{soft RG}{soft replica generation}
\newacronym{BG}{BG}{belief generation}
\newacronym{SGA}{SGA}{scalar Gaussian approximation}
\newacronym{CLT}{CLT}{central limit theorem}
\newacronym{PDF}{PDF}{probability density function}
\newacronym{QPSK}{QPSK}{quadrature phase-shift keying}
\newacronym{OQAM}{OQAM}{offset quadrature amplitude modulation}
\newacronym{LMMSE}{LMMSE}{linear minimum mean square error}
\newacronym{SNR}{SNR}{signal-to-noise ratio}
\newacronym{OOBE}{OOBE}{out-of-band emission}
\newacronym{PAPR}{PAPR}{peak-to-average power ratio}
\newacronym{AFBM}{AFBM}{Affine Filter Bank Modulation}
\newacronym{FBMC}{FBMC}{Filter Bank MultiCarrier modulation}
\newacronym{PPN}{PPN}{polyphase network}
\newacronym{SIR}{SIR}{signal-to-interference ratio}
\newacronym{AF}{AF}{ambiguity function}
\newacronym{PDA}{PDA}{probabilistic data association}
\newacronym{SBL}{SBL}{sparse Bayesian learning}
\newacronym{VGA}{VGA}{vector Gaussian approximation}
\newacronym{KL}{KL}{Kullback-Leibler}
\newacronym{GAMP}{GAMP}{generalized approximate message passing}
\newacronym{EP}{EP}{expectation propagation}
\newacronym{5G}{5G}{fifth generation}
\newacronym{4G}{4G}{fourth generation}

\newcommand\scalemath[2]{\scalebox{#1}{\mbox{\ensuremath{\displaystyle #2}}}}

\newcommand{\herm}[0]{^{\mathsf{H}}}

\begin{document}

\pagenumbering{gobble}

% \title{Affine Filter Bank Modulation: Waveform Fundamentals and ISAC Transceiver Design}
% \title{Integrated Sensing and Communications Transceiver Design for Affine Filter Bank Modulation}
% \title{Affine Filter Bank Modulation (AFBM) \\ for Integrated Sensing and Communications}
% \title{Affine Filter Bank Modulation (AFBM): A New Waveform for 6G ISAC with Low PAPR and OOBE}
% \title{Affine Filter Bank Modulation (AFBM):\\ A Low PAPR and OOBE Waveform for 6G ISAC}
% \title{Affine Filter Bank Modulation (AFBM):\\ Achieving Low PAPR and OOBE in 6G ISAC}
% \title{Affine Filter Bank Modulation (AFBM):\\ Achieving Low PAPR and OOBE in 6G ISAC}
% \title{Detection Domain Schemes \\ for Affine Filter Bank Modulation}
\title{SIR Analysis for Affine Filter Bank Modulation}

%
% \title{Affine Filter Bank Modulation for 6G ISAC: \\ Novel Waveform with Low PAPR and Low OOBE}

\author{\IEEEauthorblockN{Henrique L. Senger\IEEEauthorrefmark{1}, Gustavo P. Gonçalves\IEEEauthorrefmark{1},  Bruno S. Chang\IEEEauthorrefmark{1},  
Hyeon Seok Rou\IEEEauthorrefmark{2}, \\  Kuranage Roche Rayan Ranasinghe\IEEEauthorrefmark{2}, 
Giuseppe Thadeu Freitas de Abreu\IEEEauthorrefmark{2} and Didier Le Ruyet\IEEEauthorrefmark{3}} \vspace{1ex}

\IEEEauthorblockA{
\IEEEauthorrefmark{1}CPGEI/Electronics Department, Federal University of Technology - Paraná, Curitiba, Brazil \\[0.15ex]
% Emails: bschang@utfpr.edu.br \\[0.75ex]
\IEEEauthorrefmark{2}School of Computer Science and Engineering, Constructor University, Bremen, Germany \\[0.15ex]
%Emails: [hrou, kranasinghe, gabreu]@constructor.university\\[0.75ex]
\IEEEauthorrefmark{3}CEDRIC/Conservatoire National des Arts et Métiers - Paris, France \\
%Email: didier.le$\_$ruyet@cnam.fr
}
\vspace{-4ex}
}

\maketitle

\begin{abstract} 
    The \ac{SIR} of the \ac{AFBM} waveform is analyzed under \ac{MMSE} equalization in two domains; namely, the affine domain and the filtered \ac{TD}.
    Due to the incorporation of the \ac{DAFT} and despreading/mapping, an interesting and counter-intuitive cancellation of the unwanted combination of the channel induced interference with the orthogonality approximation error is seen in the filtered \ac{TD}, a process which does not occur in the affine domain. 
    The direct impact on \ac{BER} provides a thorough validation of the proposed analysis and explains the substantial gains in performance of the filtered \ac{TD} detection scheme as opposed to its affine domain equivalent.
\end{abstract}

\IEEEpeerreviewmaketitle

\glsresetall

\section{Introduction}

The evolution towards sixth-generation (6G) wireless networks is driven by increasingly demanding requirements that go beyond the capabilities of  increasing bandwidth of current 5G systems. 6G is envisioned to support future applications that occur in challenging scenarios like \ac{V2V}, \ac{V2X} and \ac{UAV} communications. These novel applications scenarios are characterized by severe doubly-dispersive channels with both significant Doppler spread and delay spread, that expose the limitations of existing waveforms and motivate the search for innovative physical layer solutions.

\ac{AFDM}~\cite{rou2025affine} is a recent and promising alternative waveform, providing full diversity in doubly-dispersive channels through the use of chirp based subcarrier modulation. However, \ac{AFDM} suffers from the same high \ac{PAPR} and poor spectral containment of \ac{OFDM}.
To address these limitations, we have proposed the \ac{AFBM} waveform, that integrates a pruned \ac{DAFT} precoding with an affine filterbank structure inspired by \ac{FBMC}  schemes. This approach allowed us to achieve \ac{PAPR} levels comparable to DFT-s-OFDM and \ac{OOBE} suppression similar to \ac{FBMC} systems~\cite{senger2025affinefilterbankmodulation,ranasinghe2025affinefilterbankmodulation}, while maintaining quasi-complex orthogonality (as is the case in pruned DFT spread FBMC-based systems~\cite{nissel2018pruned}) and allowing a deterministic shift of the diagonals in the affine domain channel response based on the path delay and Doppler shift indices like in \ac{AFDM}, which is the key factor enabling the robustness of the waveform by mitigating inter-path interference, in addition to the highly beneficial implications to \ac{ISAC} parameter estimation.
Yet a critical research question remains to be explored: what is the optimal detection domain for an \ac{AFBM} receiver design. Unlike conventional \ac{AFDM}, where detection is done on the discrete affine Fourier domain, \ac{AFBM}'s unique structure that combines the \ac{FBMC}
filterbank processing with an affine-domain precoding introduces additional complexity that may turn the traditional AFDM detection non-optimal. 

Time domain detection, as considered in~\cite{das2020time,wen2022downlink,liu2025entropy,shen2025timedomain}, was seen to be a good option in other delay-Doppler waveforms. 
Inspired by this and taking into account the unique structure of the \ac{AFBM} waveform, the main contribution of this work is the exploration of the effects of the detection in different domains, such as the affine and filtered-time domain, to evaluate how they affect the detection of this waveform. To do so, we will derive SIR expressions both for the \ac{AFBM} waveform itself and for a end-to-end scenario with the fading channel and the equalizer in different detection domains. 
It was found out that the inherent waveform interference cancellation properties of filtered time domain detection make it the best choice between the considered domains for detection in the \ac{AFBM} waveform.

The remainder of this paper is organized as follows. Section~\ref{sec:SysModel} of this paper presents the \ac{AFBM} system model, including its whole modulation and demodulation chain. Section~\ref{sec:AFBM_SIR} contains the derivation of the SIR analysis for the proposed detection domains, giving insights on how they deal with the waveform's inherent interference. Section~\ref{sec:SimResults} provides simulation results validating the analytical findings in various system configurations. Finally, Section~\ref{sec:Conclusion} concludes the paper and discusses implications for practical \ac{AFBM} implementation.

\section{System Model}
\label{sec:SysModel}

\subsection{Transmit Signal Model}
\label{subsec:transmit_signal_model}

Let $L$ denote the number of subcarriers and $K$ the number of time indices in a filter bank-based multicarrier system with near complex orthogonality. 
To avoid inter-filter interference, half of the $L$ subcarriers are reserved as guard bands, and transmission is carried out at twice the rate of conventional \ac{OFDM}/\ac{AFDM} systems, namely, every $L/2$ samples (see \cite{Ranasinghe_ICNC2025_oversampling} for a detailed discussion on the oversampling tradeoff). 
The system is organized into blocks of $K$ symbols, each with a duration of $T/2$ seconds. 
Within each symbol, $L$ subcarriers are employed, spaced by $F$ Hz. This results in a grid comprising $L$ points in frequency, spaced by $F$ Hz, and $K$ points in time, spaced by $T/2$ seconds in the time-frequency domain. 
Accordingly, the total bandwidth is given by $B = L F$, and the total transmission interval by $K T/2$.

Let $\mathbf{x} \in \mathcal{D}^{K\frac{L}{2} \times 1}$ denote the vector of complex transmit symbols mapped onto the defined time-frequency resources, where $\mathcal{D}$ represents a modulation alphabet of size $|\mathcal{D}|$ (e.g., \ac{QAM} constellations). 

\begin{figure*}[t!]
\centering
\includegraphics[width=\textwidth]{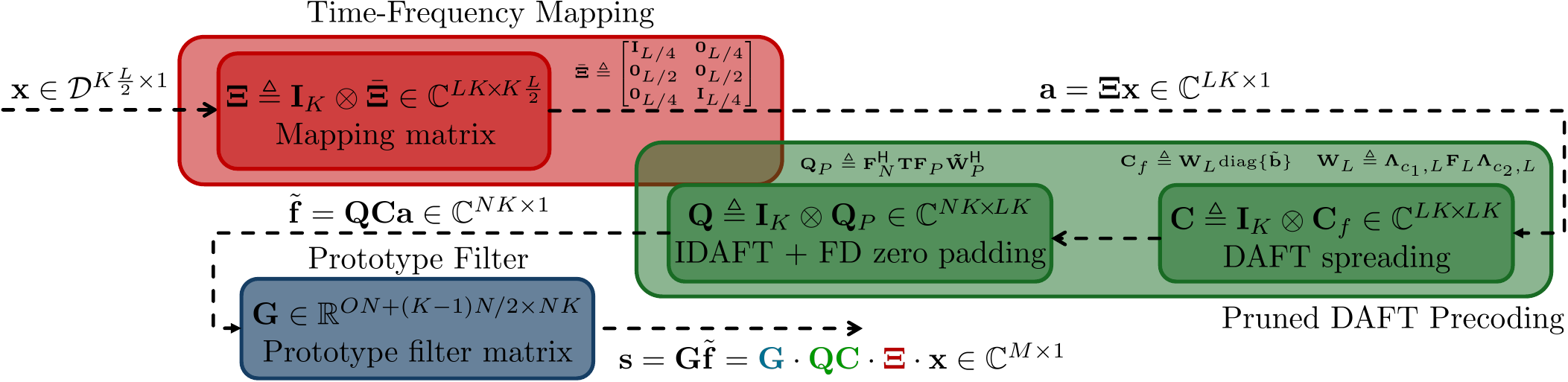}
\vspace{-3ex}
\caption{Visualization of the \ac{AFBM} modulation procedure.}
\label{fig:AFBMmod_scehamtic}
\vspace{-1ex}
\end{figure*}

The symbols in $\mathbf{x}$ are arranged in the first and last $L/4$ positions of a matrix $\mathbf{A} \in \mathbb{C}^{L \times K}$ to avoid interference from the filter bank and maximize the \ac{SIR}. This mapping is expressed as
\begin{equation}
\label{eq:positions}
\mathbf{a} \triangleq \mathrm{vec}(\mathbf{A}) = \bm{\Xi} \mathbf{x} \in \mathbb{C}^{LK \times 1},
\end{equation}
where $\mathrm{vec}(\cdot)$ denotes the column-wise vectorization operation and $\bm{\Xi} \in \mathbb{C}^{LK \times K\frac{L}{2}}$ is defined as
\begin{equation}
\label{eq:Xi}
\bm{\Xi} \triangleq \mathbf{I}_K \otimes \bar{\bm{\Xi}},
\end{equation}
with $\bar{\bm{\Xi}} \in \mathbb{C}^{L \times \frac{L}{2}}$ given by
\begin{equation}
\bar{\bm{\Xi}} \triangleq 
\begin{bmatrix}
\mathbf{I}_{L/4} & \mathbf{0}_{L/4} \\
\mathbf{0}_{L/2} & \mathbf{0}_{L/2} \\
\mathbf{0}_{L/4} & \mathbf{I}_{L/4}  
\end{bmatrix},
\end{equation}
where $\mathbf{0}_L$ denotes a zero matrix of size $L$. 
$\mathbf{A}$ is subsequently multiplied by a diagonal matrix in order to restore complex orthogonality, and then transformed into the \ac{DAFT} domain through a pruned \ac{DAFT} precoding operation. 
%
% In the following, the procedure for allowing complex orthogonality through a filter compensation stage is described, after which the pruned \ac{DAFT} precoding operation is detailed.
% \subsubsection{Restoration of Complex Orthogonality}
% \label{subsubsec:complex_orthogonality}
% A filter-bank waveform with well-localized filters and complex orthogonality can be obtained by employing the \ac{DAFT} together with a filter compensation stage that cancels interference. 
% %
% To this end, 
Let us define
\begin{equation}
\mathbf{C}_f \triangleq \mathbf{W}_{L} \mathrm{diag}\{\tilde{\mathbf{b}}\},
\label{ferf44}
\end{equation}
where $\mathbf{C}_f \in \mathbb{C}^{L \times L}$ represents the precoding process responsible for restoring complex orthogonality, and $\mathbf{W}_{L} \in \mathbb{C}^{L \times L}$ denotes the $L$-point DAFT matrix, defined as
\begin{equation}
\mathbf{W}_{L} = \mathbf{\Lambda}_{c_1,L}\mathbf{F}_{L}\mathbf{\Lambda}_{c_2,L},
\end{equation}
with
\begin{equation}
\mathbf{\Lambda}_{c_i,L} = \mathrm{diag}[e^{-j2\pi c_i (0)^2}, \dots, e^{-j2\pi c_i (L-1)^2}] \in \mathbb{C}^{L \times L}
\end{equation}
denoting an $L \times L$ diagonal chirp matrix with central digital frequency $c_i$, and where $\mathbf{F}_{L}$ denotes the normalized $L$-point \ac{DFT} matrix.

\subsubsection{Pruned DAFT Precoding}

Based on \eqref{eq:positions} and \eqref{ferf44}, the vectorized form of the DAFT-spread transmit signal $\mathbf{b} \in \mathbb{C}^{LK \times 1}$, whose matrix form is denoted by $\mathbf{B} \in \mathbb{C}^{L \times K}$ before filtering, is expressed as \vspace{-2ex}
\begin{align}
\label{eq:precodede_tx}
\mathbf{b} & \triangleq \mathrm{vec}(\mathbf{B}) = \mathrm{vec}\big(\overbrace{\mathbf{W}_L\mathrm{diag} (\tilde{\mathbf{b}})}^{\mathbf{C}_f \, \in \, \mathbb{C}^{L \times L}}\mathbf{A}\big) \\
&  = \underbrace{\big(\mathbf{I}_K \otimes \mathbf{C}_f \big)}_{\triangleq \mathbf{C} \, \in \, \mathbb{C}^{LK \times LK}} \mathrm{vec}(\mathbf{A}) =  \mathbf{C} \mathbf{a} = \mathbf{C} \bm{\Xi} \mathbf{x}, \nonumber
\end{align}
where $\otimes$ denotes the Kronecker product.

The output matrix $\mathbf{Q}_{P}$ for a given block is comprised of a IDAFT whose output is zero-padded in the frequency domain. It is obtained as
\begin{equation}
\mathbf{Q}_{P} = \mathbf{F}_{N}^{H}\mathbf{T}\mathbf{F}_P \mathbf{\tilde{W}}\herm_{P},
\label{eq:Q_P}
\end{equation}
where $\mathbf{T}\triangleq \begin{bmatrix} \mathbf{I}_{P,u} & \mathbf{0}_{P \times (N-P)} & \mathbf{I}_{P,u} \end{bmatrix}^{T}$,
with $\mathbf{T}^{T}\mathbf{T} = \mathbf{I}_P$, is an $N \times P$ matrix, with $\mathbf{I}_{P,u}\triangleq \begin{bmatrix} \mathbf{I}_{P/2} & \mathbf{0}_{P/2} \end{bmatrix}^{T}$ and $\mathbf{I}_{P,l}\triangleq \begin{bmatrix}  \mathbf{0}_{P/2} & \mathbf{I}_{P/2}  \end{bmatrix}^{T}$ 
and the pruned \ac{DAFT} $\mathbf{\tilde{W}}_{P} \in  \mathbb{C}^{L \times P}$ is defined as
\begin{equation}
\mathbf{\tilde{W}}_{P} = 
\begin{bmatrix}
\mathbf{I}_L &  \mathbf{0}_{L\times (P-L)}  
\end{bmatrix}
\mathbf{W}_P.
\label{dft_espalhada}
\end{equation}

The length $P$ of this \ac{IDAFT} at the transmitter must satisfy $L < P < N$, to ensure that the precoding stage (via \ac{DAFT}) is not nullified by the \ac{IDAFT} of the filter-bank structure, whereas $P < N$ guarantees that the chirps are sampled at a rate lower than the Nyquist rate, thereby enabling frequency containment.

The block matrix $\mathbf{Q} \in  \mathbb{C}^{NK\times LK}$ expressing the transmission of $K$ blocks is expressed as
\begin{equation}
\mathbf{Q} =  \mathbf{I}_{K} \otimes \mathbf{Q}_{P}.
\end{equation}
%
%where $\otimes$ denotes the Kronecker product, mapping $\mathbf{Q}_{P}$ into the correct time positions.

\subsubsection{Prototype Filter}

The transmitted data are obtained by convolving the precoded signal with the prototype filter impulse response through a Toeplitz filter matrix. 
Let $\mathbf{G}_p \in  \mathbb{R}^{N/2 \times N/2}$ denote the diagonal matrix of filter coefficients, i.e.,
\begin{equation}
\mathbf{G}_p = \mathrm{diag}(\mathbf{g}_p), \;\;\; p = 0,1,2,\ldots,2O-1
\end{equation}
where
\begin{equation}
\mathbf{g}_p = [g[pN/2], g[pN/2+1], \ldots , g[pN/2+N/2-1]]
\end{equation}
and $\mathbf{g}$ represents the prototype filter of length $ON$, with $O$ denoting the overlap factor. 
Accordingly, the block Toeplitz filter matrix $\mathbf{G} \in  \mathbb{R}^{ON + (K-1)N/2 \times NK}$ is defined as
\begin{equation}
\mathbf{G} = 
\scalemath{0.8}{\begin{bmatrix}
\mathbf{G}_0  & \mathbf{0} & \mathbf{0} & \mathbf{0} & \ldots  & \mathbf{0} \\
\mathbf{0} & \mathbf{G}_1  & \mathbf{G}_0  & \mathbf{0} & \ldots  & \mathbf{0} \\
\mathbf{G}_2  & \mathbf{0} & \mathbf{0} & \mathbf{G}_1  &  \ldots  & \mathbf{0} \\
\mathbf{0} & \mathbf{G}_3  & \mathbf{G}_2 & \mathbf{0} & \ldots  & \mathbf{0} \\
\vdots &  \mathbf{0} &  \mathbf{0} & \mathbf{G}_3  &  \ldots  & \mathbf{0} \\
\vdots &  \vdots &  \vdots &  \vdots &  \ddots  &   \vdots \\
\mathbf{G}_{2O-4} & \vdots &  \vdots &  \vdots & \ddots  &  \mathbf{0} \\
\mathbf{0} & \mathbf{G}_{2O-3} & \mathbf{G}_{2O-4} &  \vdots & \ddots & \mathbf{G}_1 \\
\mathbf{G}_{2O-2} &  \mathbf{0} &  \mathbf{0} &  \mathbf{G}_{2O-3} & \ddots &  \mathbf{0} \\
\mathbf{0} & \mathbf{G}_{2O-1} & \mathbf{G}_{2O-2} & \mathbf{0} & \ddots &  \mathbf{G}_3 \\
\vdots &  \mathbf{0} & \mathbf{0} & \mathbf{G}_{2O-1} & \ddots & \vdots \\
\vdots &  \vdots & \ddots &   \vdots &  \ddots &  \mathbf{0} \\
\mathbf{0} &   \mathbf{0} &  \ldots & \mathbf{0} &  \ddots & \mathbf{G}_{2O-1} 
\end{bmatrix}}.
\label{fhddg}
\end{equation}
The structure of $\mathbf{G}$, through the inclusion of $\mathbf{0}_{N/2}$ matrices, ensures that the transmitted symbols are delayed from one another every $N/2$ samples. 
This representation as a sum of delayed matrices is detailed in \cite{pereira2022generalized}. 

%\textcolor{red}{I think the rest of this section should be moved after the description of the prototype filter since $\mathbf{Q}_{P}$ and  $\widetilde{G}$ need to be introduced -   it can be a subsection "complex orthogonality" after the subsection prototype filter}

\subsubsection{Complex Orthogonality}

To preserve complex orthogonality, $\mathbf{C}_f$ must be chosen such that the following condition is satisfied:
\begin{equation}
\mathbf{C}\herm_f\mathbf{Q}_{P}^{H}\mathbf{\widetilde{G}}^T\mathbf{\widetilde{G}}\mathbf{Q}_{P} \mathbf{C}_f \approx \mathbf{U},
\label{eq:complex_orthogonality}
\end{equation}
where $\mathbf{U} \in \mathbb{R}^{L \times L}$ is a diagonal matrix with unit values in the first and last $L/4$ positions and zeros elsewhere. 

Here, $\mathbf{\widetilde{G}}\in \mathbb{R}^{ON \times N}$ denotes the filtering matrix corresponding to the transmission of a single multicarrier symbol, expressed as 
%$\mathbf{\widetilde{G}} = [\mathbf{G}_0; \mathbf{G}_1; \ldots; \mathbf{G}_{2O-1}] \in \mathbb{R}^{ON \times N}$. 
%

\begin{equation}
    \mathbf{\widetilde{G}} =
    \begin{bmatrix}
\mathbf{G}_0 & \mathbf{0}_{N/2} \\
\mathbf{0}_{N/2} & \mathbf{G}_1 \\
\mathbf{G}_2 & \mathbf{0}_{N/2}  \\
\vdots & \vdots \\
\mathbf{0}_{N/2} & \mathbf{G}_{2O-1}
\end{bmatrix},
\end{equation}

By substituting \eqref{ferf44} into \eqref{eq:complex_orthogonality}, the $\tilde{l}$-th element of $\tilde{\mathbf{b}}$ is obtained as
\begin{equation}
[\mathbf{\tilde{b}}]_{\tilde{l}} = 
\begin{cases} 
\sqrt{\frac{1}{[\mathbf{\tilde{c}}]_{\tilde{l}}}}, & \tilde{l} \in \left[ 0,\ldots,\tfrac{L}{4}-1 \right] \cup \left[ L-\tfrac{L}{4},\ldots,L-1 \right] \\[1ex]
0, & \text{otherwise},
\end{cases}
\end{equation}
with
\begin{equation}
\mathbf{\tilde{c}} \triangleq \mathrm{diag}\{\mathbf{W}\herm_L\mathbf{Q}_{P}^{H}\mathbf{\widetilde{G}}^T\mathbf{\widetilde{G}}\mathbf{Q}_{P}\mathbf{W}_L\}.
\end{equation}

The compensation stage thus comprises a multiplicative factor that cancels the interference in the transmitted symbols introduced by the filter coefficients. 
Since the coefficients are derived from a pre-defined prototype filter \cite{pereira2023two}, they are assumed to be known. 
Correct compensation is guaranteed when the interference is limited to a single coefficient, which is achieved by selecting a prototype filter overlap factor $O \leq 1.5$ \cite{pereira2022generalized}. 
If the overlap factor exceeds this threshold, off-diagonal interference appears in \eqref{eq:complex_orthogonality}, thereby reducing the achievable \ac{SIR}. 

\subsubsection{Effective Transmit Signal}

In all, the complete \ac{AFBM} transmit signal in the \ac{TD} can be expressed in terms of the precoding matrix $\mathbf{C}_f$ in \eqref{ferf44}, the modified \ac{IDAFT} matrix $\mathbf{Q}_P$ in \eqref{eq:Q_P}, and the filter matrix $\mathbf{G}$ in \eqref{fhddg}, by exploiting Kronecker product identities as
\begin{align}
\label{eq:td_tx_signal}
\mathbf{s} & = \mathbf{G} \mathbf{Q} \mathbf{C} \mathbf{a}  
= \mathbf{G} \big(\mathbf{I}_{K} \otimes \mathbf{Q}_{P}\big) \cdot \big(\mathbf{I}_K \otimes \mathbf{C}_f \big) \mathbf{a} \in \mathbb{C}^{M \times 1} \nonumber \\
& = \mathbf{G} \big(\mathbf{I}_{K} \otimes \mathbf{Q}_{P} \mathbf{C}_f \big) \bm{\Xi} \mathbf{x},
\end{align}
where $M \triangleq ON + \tfrac{N}{2}(K-1)$. 

%\textcolor{red}{Bruno: provide a filterbank-ready formulation?}

%\subsubsection{Additional Constraints}

%It follows from the above that the proposed structure constitutes an affine-precoded filter-bank scheme, which can be implemented efficiently using a \ac{PPN} together with an \ac{IFFT}. 
%
The proposed waveform can be interpreted as a modification of \ac{AFDM}, where the standard sinc-chirp subcarriers are replaced with chirp-filtered subcarriers, where the considered filter is well localized both in time and in frequency.

\vspace{-1ex}
\subsection{Receive Signal Model}
\label{subsec:received_signal}

The transmit signal vector $\mathbf{s}$ in \eqref{eq:td_tx_signal} is propagated through a time-varying multipath channel, i.e., a doubly-dispersive channel. %, concisely  represented by the circular convolutional matrix \textcolor{red}{$\mathbf{H} \in \mathbb{C}^{(K-1)M \times (K-1)M}$ should be $ \mathbf{H} \in \mathbb{C}^{M \times M}$} \cite{nissel2018pruned}. 
Consequently, the received signal $\mathbf{r} \in \mathbb{C}^{M \times 1}$ is expressed as
\begin{equation}
\mathbf{r} \triangleq \mathbf{H}
\mathbf{G} \Big(\mathbf{I}_{K} \otimes \mathbf{Q}_{P} \mathbf{C}_f\Big) \bm{\Xi} \mathbf{x} + \mathbf{n},
\end{equation}
where $\mathbf{H} \in \mathbb{C}^{M \times M}$ denotes the doubly-dispersive channel composed of $R$ resolvable paths. 
Each $r$-th path induces a delay $\tau_r \in [0, \tau^\mathrm{max}]$ and a Doppler shift $\nu_r \in [-\nu^\mathrm{max}, +\nu^\mathrm{max}]$, with normalized integer delay $\ell_r \triangleq \lfloor \tfrac{\tau_r}{T_\mathrm{s}} \rceil \in \mathbb{N}_0$ and normalized Doppler $f_r \triangleq \tfrac{N\nu_r}{f_\mathrm{s}} \in \mathbb{R}$, where $f_\mathrm{s} \triangleq \tfrac{1}{T_\mathrm{s}}$ denotes the sampling frequency. 
The noise vector $\mathbf{n} \in \mathbb{C}^{M \times 1}$ represents \ac{AWGN} samples with variance $\sigma_n^2$. 

% For simplicity, it is assumed that the doubly-dispersive channel remains constant during the $K$ time slots.\footnote{This assumption allows the definition $\bar{N} = NK$ for the linear detection procedure in Section~\ref{secReceivercomm}. A time-varying case can also be considered but is left for future work.} 

%As described in \cite{Rou_SPM_2024}, the channel matrix $\mathbf{H}$ is expressed as  \vspace{-1ex}
%
%\begin{equation}
%\mathbf{H} \triangleq \sum_{r=1}^{R} h_r \mathbf{\Phi}_{r} \mathbf{Z}^{f_r} \mathbf{\Pi}^{\ell_r} \in \mathbb{C}^{M \times M},
%\vspace{-1ex}
%\end{equation}
%
%where $h_r \in \mathbb{C}$ denotes the complex fading coefficient of the $r$-th path, $\mathbf{\Phi}_r \in \mathbb{C}^{M \times M}$ is the IDAFT-based chirp-cyclic prefix phase matrix, $\mathbf{Z} \in \mathbb{C}^{M \times M}$ is the diagonal roots-of-unity matrix, and $\mathbf{\Pi} \in \mathbb{C}^{M \times M}$ is the circular left-shift matrix. 
%
%The phase matrix is given by
%
%\begin{equation}
%\mathbf{\Phi}_{r} \triangleq \mathrm{diag}\big[e^{-j2\pi \cdot \phi(\ell_r)}, \ldots, e^{-j2\pi \cdot \phi(1)}, 1, \ldots, 1\big],
%\label{eq:CCP_phase_matrix}
%\end{equation}
%
%with $\phi(m) \triangleq c_1(M^2 - 2Mm)$. 
%
%The matrix $\mathbf{Z}$ is defined as
%
%\begin{equation}
%\mathbf{Z} \triangleq \mathrm{diag}\big[e^{-j2\pi \tfrac{0}{M}}, \ldots, e^{-j2\pi \tfrac{M-1}{M}}\big],
%\label{eq:Z_matrix}
%\end{equation}
%
%and is raised to the power $f_r$. 
%

As described in \cite{raviteja2018practical}, the channel matrix  $\mathbf{H}$ is expressed as  \vspace{-1ex}
\begin{equation}
\mathbf{H} \triangleq \sum_{r=1}^{R} h_r  \mathbf{Z}^{f_r} \mathbf{\Pi}^{\ell_r} \in \mathbb{C}^{M \times M},
\vspace{-1ex}
\end{equation}
where $h_r \in \mathbb{C}$ denotes the complex fading coefficient of the $r$-th path, $\mathbf{Z} \in \mathbb{C}^{M \times M}$ is the diagonal roots-of-unity matrix, and $\mathbf{\Pi} \in \mathbb{C}^{M \times M}$ is the circular shift matrix obtained by left shifting the $M \times M$ identity matrix once \cite{Rou_SPM_2024}. 
The matrix $\mathbf{Z}$ is defined as
\begin{equation}
\mathbf{Z} \triangleq \mathrm{diag}\big[e^{-j2\pi \tfrac{0}{M}}, \ldots, e^{-j2\pi \tfrac{M-1}{M}}\big],
\label{eq:Z_matrix}
\end{equation}
and is raised to the power $f_r$.
In total, $\mathbf{H}$ is formed by $R$ diagonals, with positions determined by the path delays and coefficients modulated by the Doppler shifts.

The received signal is next demodulated by $(\mathbf{G}\mathbf{Q})\herm$, producing $\mathbf{\tilde{a}} \in \mathbb{C}^{LK \times 1}$. 
Considering all time slots, the detected symbols $\mathbf{\tilde{B}} \in \mathbb{C}^{L \times K}$ are obtained through the \ac{IDAFT} combined with the compensation stage as
\begin{equation}
\mathbf{\tilde{B}} = \mathbf{W}\herm_L \mathrm{diag}\{\tilde{\mathbf{b}}\} \mathbf{\tilde{A}},
\label{12ghtj}
\end{equation}
where $\mathbf{\tilde{A}} = [\mathbf{\tilde{a}}_0;\mathbf{\tilde{a}}_1;\ldots;\mathbf{\tilde{a}}_{K-1}] \in \mathbb{C}^{L \times K}$. 
Since no data are transmitted in the intermediate $L/2$ positions of $\mathbf{A}$ (as detailed in~\ref{eq:Xi}), these are discarded at the receiver. 
For symmetry, the compensation stage is applied at both transmitter and receiver, but it is effective only on one side. 

Concatenating all effects, the final received signal $\mathbf{y} \in \mathbb{C}^{K\frac{L}{2} \times 1}$ in the absence of noise is expressed as
\begin{equation}
\label{eq:final_IO}
\mathbf{y} \triangleq \bm{\Xi}\herm \Big(\mathbf{I}_{K} \otimes \mathbf{C}_f\herm \mathbf{Q}_{P}\herm \Big) \mathbf{G}\herm \mathbf{H} \mathbf{G} \Big(\mathbf{I}_{K} \otimes \mathbf{Q}_{P} \mathbf{C}_f\Big) \bm{\Xi} \mathbf{x},
\end{equation}
where the effective channel matrix in the affine domain is defined as
\begin{equation}
\mathbf{H}_\text{AFB} \triangleq  \Big(\mathbf{I}_{K} \otimes \mathbf{C}_f\herm \mathbf{Q}_{P}\herm \Big) \mathbf{G}\herm \mathbf{H} \mathbf{G} \Big(\mathbf{I}_{K} \otimes \mathbf{Q}_{P} \mathbf{C}_f\Big),
\label{eq:Heff_afb}
\end{equation}
and the filtered time domain effective channel matrix $\bar{\mathbf{H}} \in \mathbb{C}^{NK \times K\frac{L}{2}}$ is given by
\begin{equation}
    \bar{\mathbf{H}} = \mathbf{G}\herm \mathbf{H} \mathbf{G} \Big(\mathbf{I}_{K} \otimes \mathbf{Q}_{P} \mathbf{C}_f\Big)\bm{\Xi}.
    \label{eq:Heff_ftd}
\end{equation}

\section{SIR Analysis}
\label{sec:AFBM_SIR}

We recall that the AFBM waveform does not have full complex orthogonality, as seen in~\eqref{eq:complex_orthogonality}.
The residual interference of the waveform can be quantified by the SIR, which is expressed as
\begin{equation}
    \text{SIR}_{\mathbf{W}} = \frac{\frac{LK}{2}}{|| \bm{\Xi}\herm \Big(\mathbf{I}_{K} \otimes \mathbf{C}_f\herm \mathbf{Q}_{P}\herm \Big) 
\mathbf{G}\herm  \mathbf{G} \Big(\mathbf{I}_{K} \otimes \mathbf{Q}_{P} \mathbf{C}_f\Big) \bm{\Xi} ||^2 - \frac{LK}{2}}.
    \label{eq:sir}
\end{equation}
Several parameters determine the SIR in this waveform, such as the choice of the prototype filter and the values of $K$, $L$, $N$ and $P$. 
As seen in~\cite{nissel2018pruned,pereira2021novel}, the optimum SIR values of pruned DFT precoded \ac{FBMC} waveforms will be obtained with large values of $L$ and $N$ and a prototype filter with $O \leq 1.5$, due to the inherent structure of the waveform. 
However, $P$ is exclusive to the AFBM waveform - we recall that this parameter controls spectral occupation and must be smaller than $N$.
%\textcolor{red}{Bruno: and c2?} %With the truncated Hermite filter as the prototype one and $O = 1.5$ the obtained SIR is taltal dB, while when choosing the PHYDYAS one with $O = 4$ the obtained SIR is taltal dB. 

Let us now analyze the end-to-end scenario, considering the fading channel and the equalizer.
In the following, we will limit ourselves to high-complexity MMSE equalization to simplify the analysis. However, the insights obtained by this analysis can be applied to the choice of detection domain using other structures, such as Gaussian belief propagation or message passing algorithms. The received signal in the affine domain after the DAFT in the receiver is
\begin{equation}
\mathbf{r} = \Big(\mathbf{I}_{K} \otimes \mathbf{C}_f\herm \mathbf{Q}_{P}\herm \Big) 
\mathbf{G}\herm \mathbf{H}
\mathbf{G} \Big(\mathbf{I}_{K} \otimes \mathbf{Q}_{P} \mathbf{C}_f\Big) \bm{\Xi} \mathbf{x} + \mathbf{n}_{\text{AFB}},
\end{equation}
whereas the estimated signal using affine-domain detection is 
\begin{align}
\begin{matrix}  \mathbf{\tilde{x}}_{\text{AFB}} &=  \underbrace{ \bm{\Xi}\herm \mathbf{E}_\text{AFB} \Big(\mathbf{I}_{K} \otimes \mathbf{C}_f\herm \mathbf{Q}_{P}\herm \Big) 
\mathbf{G}\herm \mathbf{H}
\mathbf{G} \Big(\mathbf{I}_{K} \otimes \mathbf{Q}_{P} \mathbf{C}_f\Big) \bm{\Xi}}_{\mathbf{\Delta_{\text{AFB}}}}\mathbf{a} + \\ & + \bm{\Xi}\herm \mathbf{E}_\text{AFB}\mathbf{n}_{\text{AFB}},  \end{matrix} 
\label{eq:estsig_afdetection}
\end{align}
with $\Delta_{\text{AFB}} \in \mathbb{C}^{ \frac{LK}{2} \times \frac{LK}{2}}$. The affine-domain MMSE equalizer can be given by
\begin{equation}
\label{eq:LMMSE_AFB}
\mathbf{E}_\text{AFB} = \left( \mathbf{H}_\text{AFB}\herm \mathbf{H}_\text{AFB} + \sigma^2_n \mathbf{I}_{K\frac{L}{2}} \right)^{-1} \mathbf{H}_\text{AFB}\herm.
\end{equation}

If detection in the filtered TD detection is considered, the received signal after filtering is 
\begin{equation}
\bar{\mathbf{r}} = \bar{\mathbf{H}}  \mathbf{x} + \bar{\mathbf{n}}.
\end{equation}
whereas the estimated signal using filtered TD detection is 
\begin{eqnarray}
\begin{matrix}  \mathbf{\tilde{x}}_{\text{FTD}} = \underbrace{ \mathbf{E}_\text{FTD}
\mathbf{G}\herm \mathbf{H}
\mathbf{G} \Big(\mathbf{I}_{K} \otimes \mathbf{Q}_{P} \mathbf{C}_f\Big) \bm{\Xi}}_{\mathbf{\Delta_{\text{FTD}}}}\mathbf{a} + \mathbf{E}_\text{FTD}\mathbf{\bar{n}},  \end{matrix} 
\label{eq:estsig_htddetection}
\end{eqnarray}
where $\Delta_{\text{FTD}} \in \mathbb{C}^{ \frac{LK}{2} \times \frac{LK}{2}}$ and the filtered time domain MMSE equalizer can be given by
\begin{equation}
\label{eq:HTD_LMMSE}
\mathbf{E}_\text{FTD} = \left( \bar{\mathbf{H}}\herm \bar{\mathbf{H}} + \sigma^2_n \mathbf{I}_{K\frac{L}{2}} \right)^{-1} \bar{\mathbf{H}}\herm.
\end{equation}

The interference taking into account the channel is described by the off-diagonal elements of $\mathbf{\Delta}_{\text{d}}$, with $\text{d} \in \text{AFD}, \text{FTD}$ representing the chosen detection domain. The SIR conditioned on a given channel realization $\mathbf{H}$ can be thus calculated by
\begin{eqnarray}
\text{SIR}_{\mathbf{H}} = \frac{\frac{LK}{2}}{|| \Delta_{\text{d}}||^{2} - \frac{LK}{2}    }.
\label{eq:sir_channel}
\end{eqnarray}

\eqref{eq:sir_channel} includes the combination of the channel induced interference with the orthogonality approximation error. 
An interesting insight appears by analyzing~\eqref{eq:estsig_afdetection} and~\eqref{eq:estsig_htddetection}. Since the equalizer in the filtered time domain incorporates the DAFT and despreading/mapping, it is able to cancel the undesired combination of the channel induced interference with the orthogonality approximation error and improve the SIR, a process which does not happen in the affine domain detection. 
%

%

% A link here with the effective channel matrix in both domains discussing sparsity

\section{Simulation Results}
\label{sec:SimResults}

% Simulation parameters here
For the simulations, the total number of subcarriers $L$ is 128 and the filter bank size $N$ is 256.
Each transmission was composed of $K = 8$ symbols, with a carrier frequency $f_c$ of 4 GHz.
The channel is a doubly dispersive one, with three resolvable paths and corresponding normalized delays and digital Doppler shifts.
We recall that the chirp frequencies for each (I)DAFT were chosen to uphold the orthogonality condition \cite{Rou_SPM_2024} $2(f^{\text{max}} + \xi)(\ell^{\text{max}}+1) + \ell^{\text{max}} \leq P$, where  $f^{\text{max}}$ and $\ell^{\text{max}}$ are, respectively, the maximum normalized digital Doppler shift and delay of the channel and $\xi \in \mathbb{N}_0$ is a free parameter determining the so-called guard width, denoting the number of additional guard elements around the diagonals to anticipate for Doppler-domain interference.
In the considered case, the channel parameters were chosen as $\ell^{\text{max}} = 16$ with randomly chosen integer values and $f^{\text{max}} = 2$ with randomly chosen fractional values. The following values were selected for the $c_2$ parameter - $c_{2,L} = \tfrac{1}{\pi L^2}$ and $c_{2,P} = \tfrac{1}{\pi P^2}$ - with these parameters, a low PAPR is obtained in AFBM~\cite{ranasinghe2025affinefilterbankmodulation}.

\begin{figure}[!h]
\centering
\includegraphics[width=1\columnwidth]{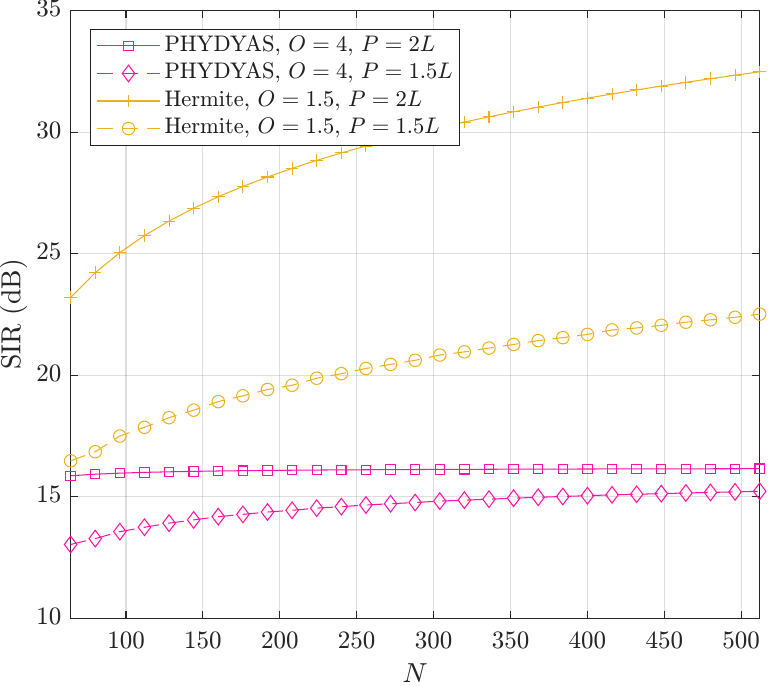}%
\vspace{-1ex}
\caption{\ac{SIR} of the \ac{AFBM} waveform with the Hermite ($O=1.5$) and PHYDYAS ($O=4$) prototype filters for different values of $L,P$ and $N$.}
\label{fig:sirafbm}
% \vspace{-3ex}
\end{figure}

%\textcolor{red}{Bruno: SIR variation with the values of c2}
Figure~\ref{fig:sirafbm} presents an analysis of the SIR of the AFBM waveform, according to~\eqref{eq:sir}.
It can be seen that the SIR value is at its highest when $N = P$ and the Hermite prototype filter with $O = 1.5$ is chosen. If $P$ is reduced and the Hermite filter is replaced with the PHYDYAS one there is an improvement in spectral containment and in \ac{OOBE}, as seen in~\cite{senger2025affinefilterbankmodulation}. However, as a consequence the SIR is reduced. As an example, with the PHYDYAS filter, $O=4$, $L = 64$, $P = 96$ and $N=128$ the SIR is around 15 dB, which is a very low value for practical scenarios.

\begin{figure*}[b]    
	\centering
	
	\begin{subfigure}[b]{0.235\textwidth}
		\centering
		\includegraphics[width=\textwidth]{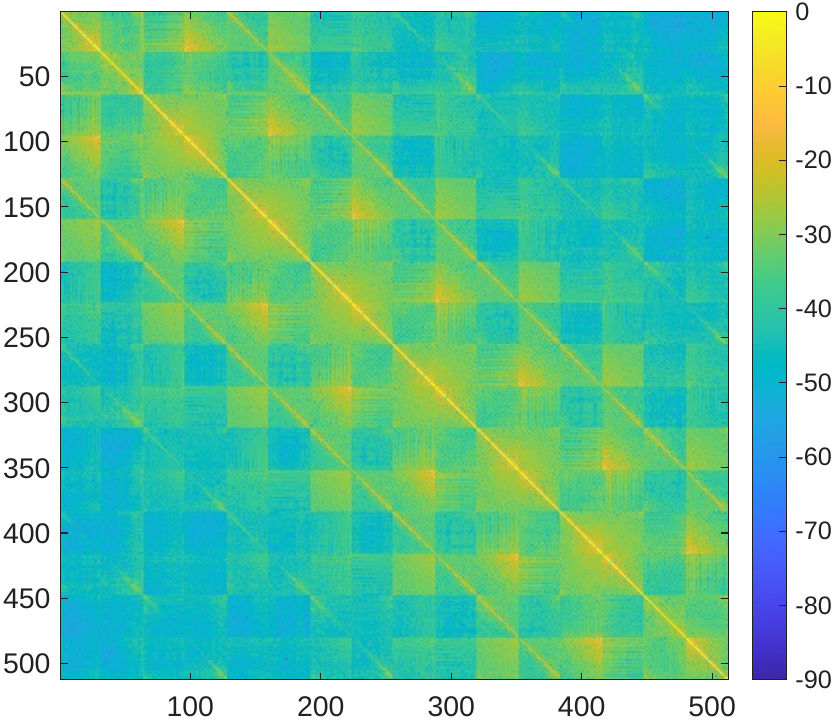}
		\caption{$P=192$, SIR = 14.87 dB}
		\label{fig:sir_affine_hermite_P96N128}
	\end{subfigure}
	\hfill
	\begin{subfigure}[b]{0.235\textwidth}
		\centering
		\includegraphics[width=\textwidth]{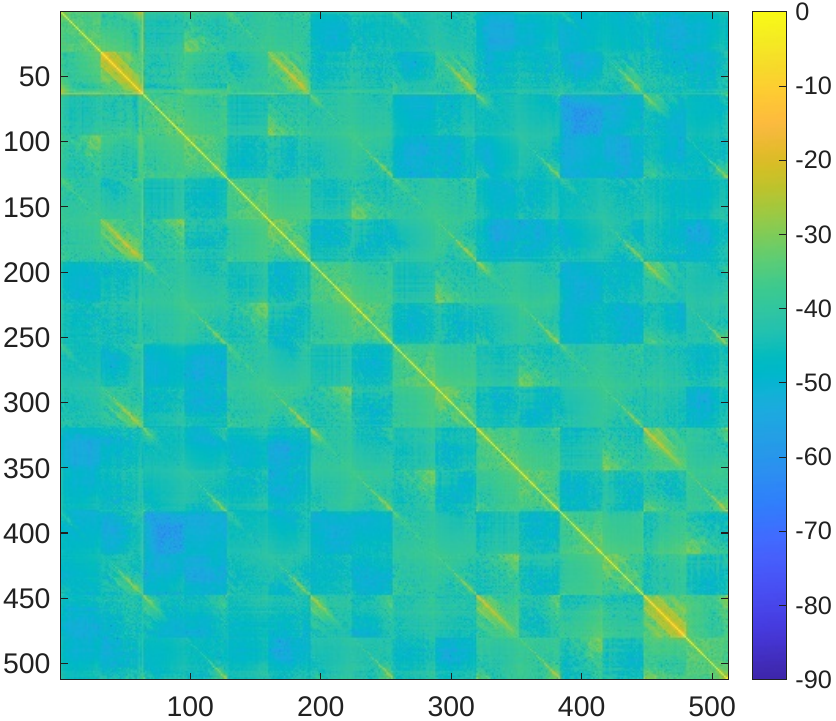}
		\caption{$P=256$, SIR = 20.67 dB }
		\label{fig:sir_affine_hermite_P128N128}
	\end{subfigure}
	\hfill
	\begin{subfigure}[b]{0.235\textwidth}
		\centering
		\includegraphics[width=\textwidth]{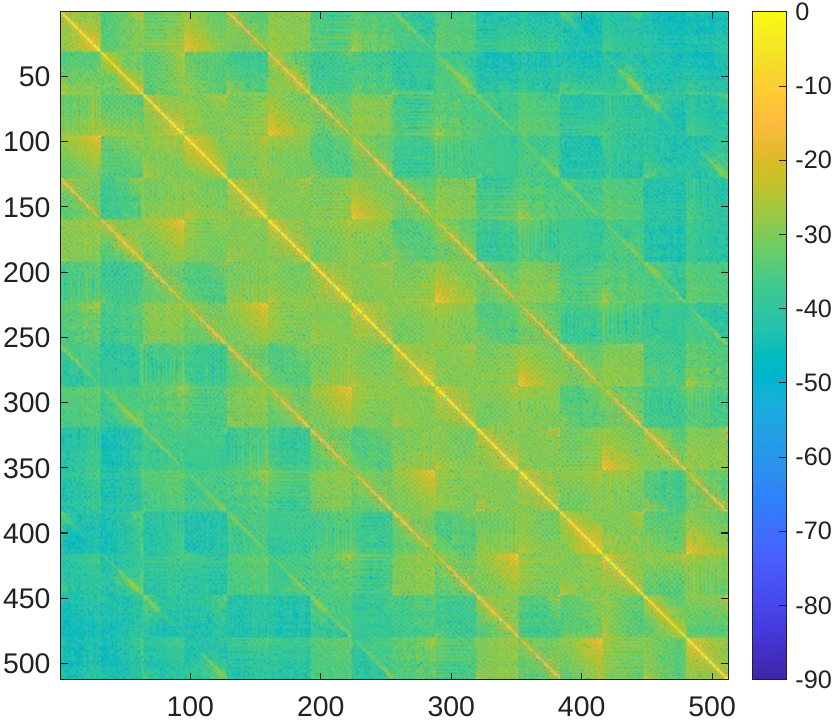}
		\caption{$P=192$, SIR = 12.34 dB}
		\label{fig:sir_affine_phydyas_P96N128}
	\end{subfigure}
	\hfill
	\begin{subfigure}[b]{0.235\textwidth}
		\centering
		\includegraphics[width=\textwidth]{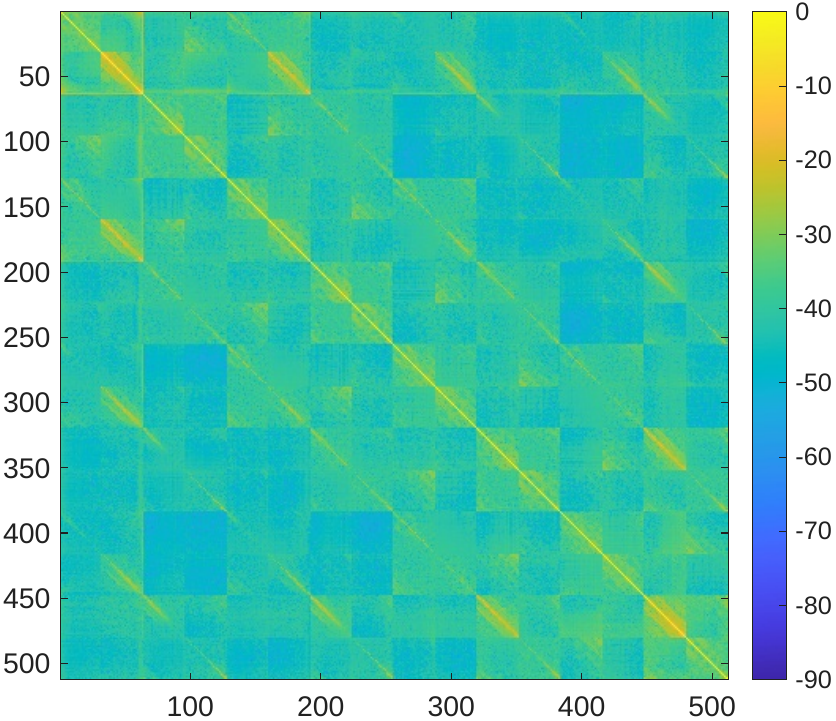}
		\caption{$P=256$, SIR = 20.08 dB}
		\label{fig:sir_affine_phydyas_P128N128}
	\end{subfigure}     
	\caption{Desired signal, off-main diagonal interference and SIR with affine detection for $N = 256$. (a) and (b) use the Hermite filter, with $O=1.5$, while (c) and (d) use the PHYDYAS filter, with $O=4$.}
	\label{fig:sir_affine}
\end{figure*}

\begin{figure*}[b]    
	\centering
	\begin{subfigure}[b]{0.235\textwidth}
		\centering
		\includegraphics[width=\textwidth]{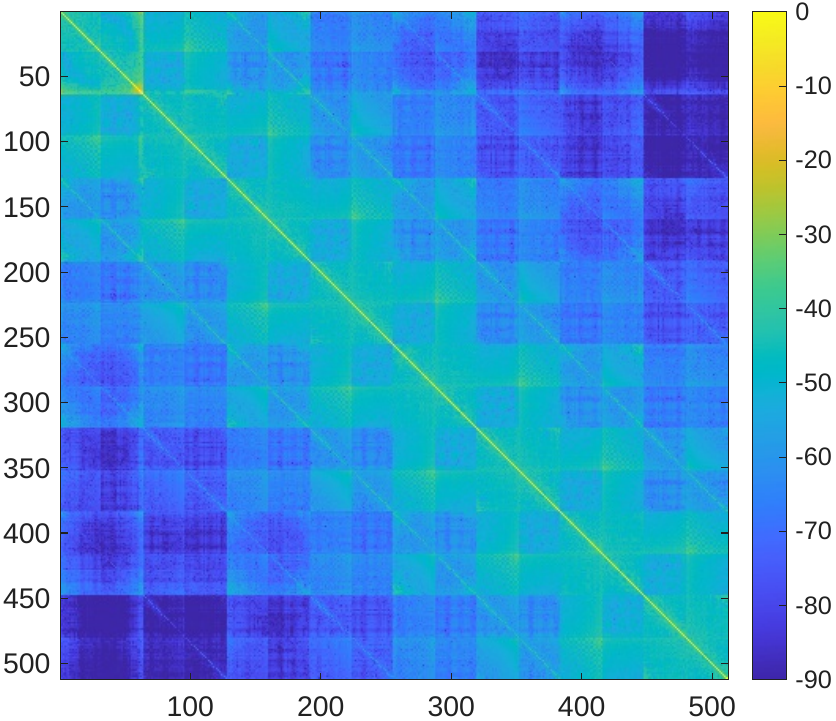}
		\caption{$P=192$, SIR = 43.01 dB} 
		\label{fig:sir_hybrid_hermite_P256N256}
	\end{subfigure}
	\hfill
	\begin{subfigure}[b]{0.235\textwidth}
		\centering
		\includegraphics[width=\textwidth]{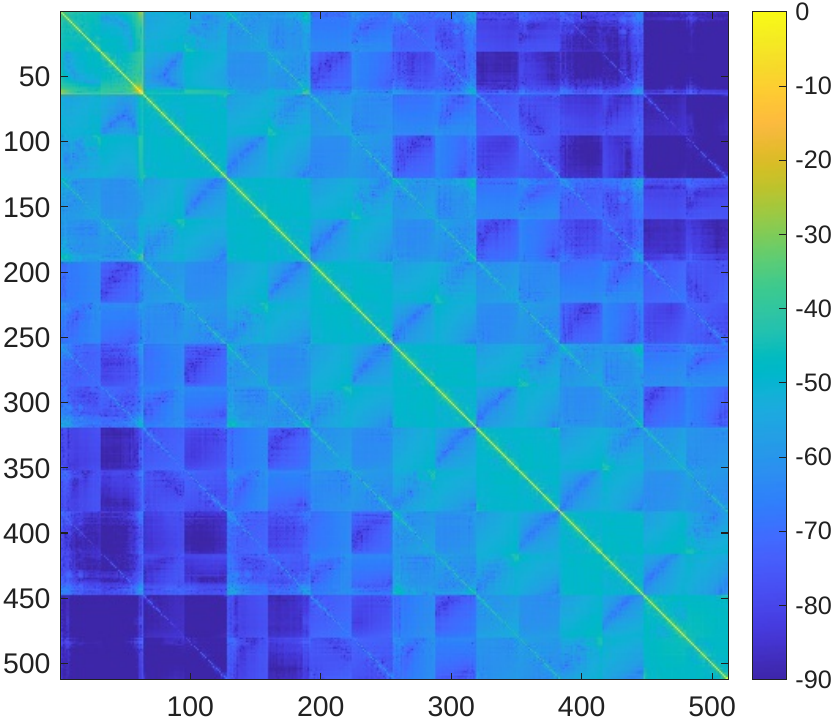}
		\caption{$P=256$, SIR = 45.18 dB} 
		\label{fig:sir_hybrid_hermite_P192N256}
	\end{subfigure}
	\hfill
	\begin{subfigure}[b]{0.235\textwidth}
		\centering
		\includegraphics[width=\textwidth]{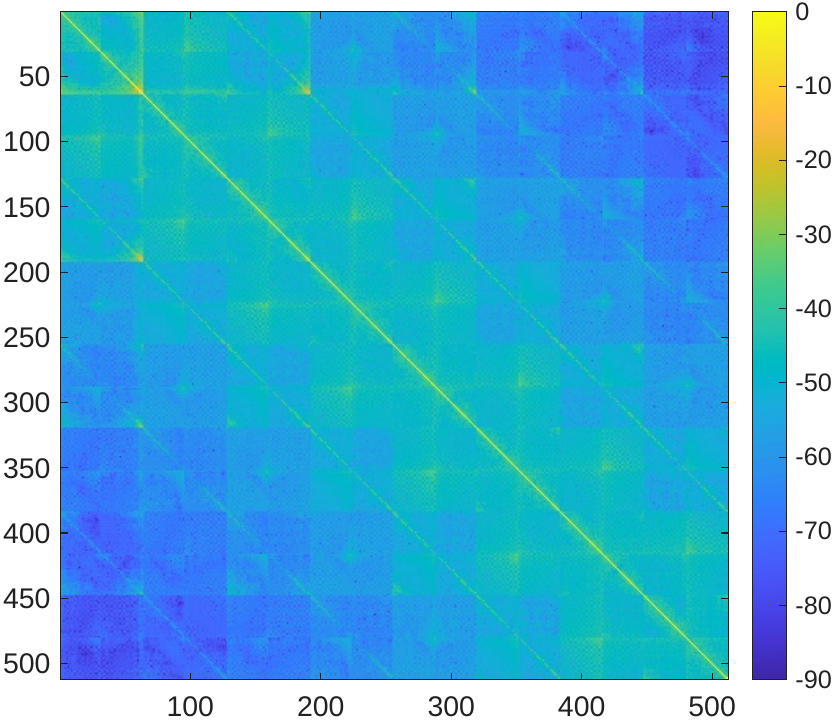}
		\caption{$P=192$, SIR = 42.43 dB} 
		\label{fig:sir_hybrid_phydyas_P128N128}
	\end{subfigure}
	\hfill
	\begin{subfigure}[b]{0.235\textwidth}
		\centering
		\includegraphics[width=\textwidth]{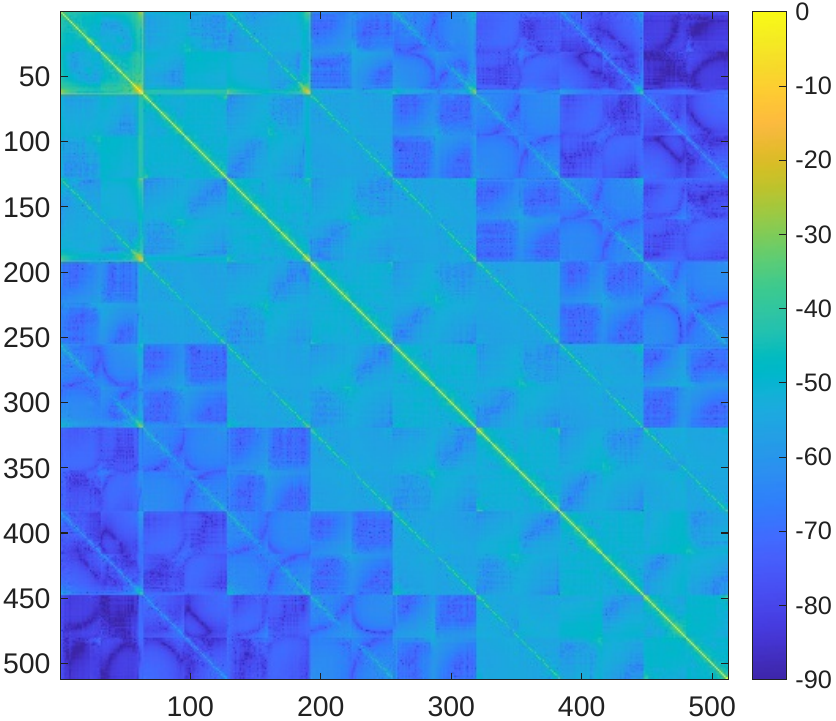}
		\caption{$P=256$, SIR = 43.44 dB} 
		\label{fig:sir_hybrid_phydyas_P96N128}
	\end{subfigure}
	\caption{Desired signal, off-main diagonal interference and SIR with filtered time domain detection for $N = 256$. (a) and (b) use the Hermite filter, with $O=1.5$, while (c) and (d) use the PHYDYAS filter, with $O=4$.}
	\label{fig:sir_hybrid}
\end{figure*}

Now taking into account the channel and the equalizer, the desired signal, together with the residual off-main diagonal interference in each position in the estimated signal in all of the transmitted blocks, can be seen in Figures~\ref{fig:sir_affine} and~\ref{fig:sir_hybrid}. The corresponding SIR values were then calculated with~\eqref{eq:sir_channel}. We recall that the presented interference/SIR values were averaged over 200 channel realizations. While when using affine domain detection there is significant residual off-main diagonal interference (especially in the scenarios with the PHYDYAS prototype filter), the same cannot be said with respect to the case using filtered TD detection. 
This is due to the filtered TD equalizer dealing the waveform interference together with the fading channel. Thus, these lower residual interference when using hybrid time domain leads to a higher overall SIR value. 

Providing a greater insight, Table~\ref{tab:sir} presents average, maximum and minimum SIR values for the considered detection domains using the Hermite (with $O=1.5$) and PHYDYAS (with $O=4$) filters and $P = 192, 256$ over all of the previously considered channel realizations. It can be seen that even the worst case when using filtered time domain detection has a $\sim 5$ dB advantage with respect to the best average in the affine domain. Moreover, the minimum SIR in the worst scenario using filtered time domain detection is close than the the waveform SIR obtained in the best scenario, demonstrating the interference cancellation properties of detection in this domain. Finally, considering a scenario with good \ac{OOBE} characteristics and spectral efficiency ($P=192$, PHYDYAS prototype filter) there is a $\sim$ 30 dB advantage in average SIR when considering filtered time domain detection instead of the affine domain.

\begin{table}[!hbt]
\caption{SIR values in dB for different detection domains for $L = 128$, $N=256$ and the Hermite and PHYDYAS prototype filters.}
\label{tab:sir}
\begin{center}
\begin{tabular}{p{4.2cm}p{1cm}p{1cm}p{1cm}}
\toprule
Scenario & Average & Maximum & Minimum \\
\midrule
Affine, Hermite, $P=192$ & 14.87 & 15.90 & 12.31 \\
Affine, Hermite, $P=256$ & 20.67 & 47.04 & 14.51 \\
Affine, PHYDYAS, $P=192$ & 12.34 & 13.12 & 10.43 \\
Affine, PHYDYAS, $P=256$ & 20.08 & 47.31 & 14.02 \\
Filtered time, Hermite, $P=192$ & 43.01 & 71.98 & 28.73 \\
Filtered time, Hermite, $P=256$ & 45.18 & 68.66 & 27.95 \\
Filtered time, PHYDYAS, $P=192$ & 42.43 & 58.89 & 28.36 \\
Filtered time, PHYDYAS, $P=256$ & 43.44 & 70.46 & 25.88 \\
\bottomrule
\end{tabular}
\end{center}
\end{table}

\begin{figure}[H]
\centering
\includegraphics[width=1.0\columnwidth]{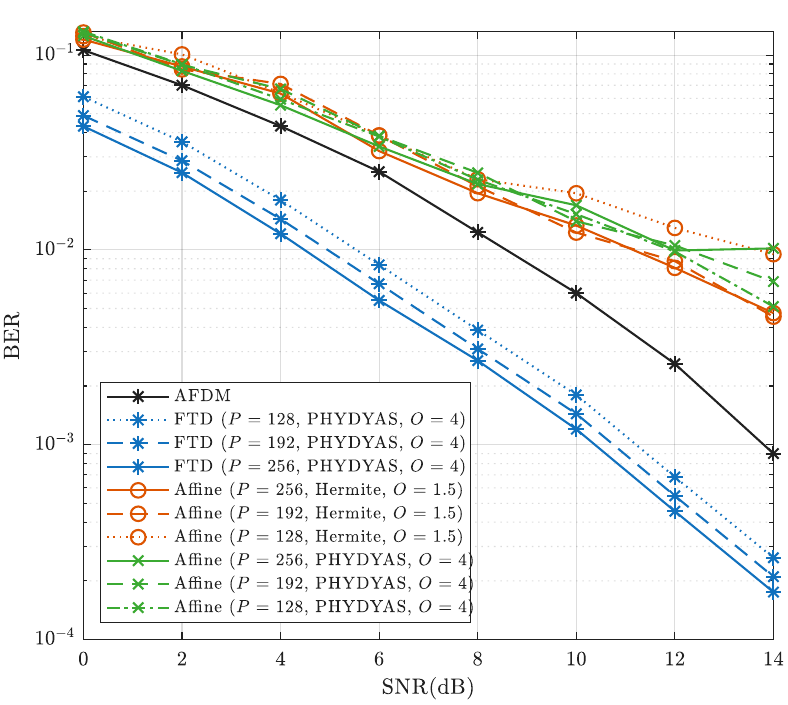}%
\vspace{-1ex}
\caption{\ac{BER} of the \ac{AFBM} waveform with the Hermite and PHYDYAS prototype filters for different values of $L,P$ and $N$ using different detection domains.}
\label{fig:berafbm}
% \vspace{-3ex}
\end{figure}

Finally, Figure~\ref{fig:berafbm} shows \ac{BER} results for all of the aforementioned detection domains and for different values of $P$ . %\textcolor{red}{Bruno: and c2?}
The error results validate the SIR analysis presented above, with the systems using filtered time detection having significantly better error performance (a $\sim$5 dB advantage even at a low BER of $10^{-2}$) and near-insensitivity to the choice of $P$ and the prototype filter.
This allows the usage of better localized prototype filters and an increase in spectral efficiency in AFBM without the inherent waveform SIR penalty.

%\textcolor{red}{Bruno: Gram matrices for different parameters and domains to link to the next part?}

\section{Conclusion}
\label{sec:Conclusion}

In this work, we have presented an \ac{SIR} analysis of the \ac{AFBM} for different parameters and detection domains. It was found out that the \ac{SIR} of the waveform is reduced when optimizing its \ac{OOBE} characteristics and spectral efficiency. This is due to the extra interference generated by the bandwidth occupation reduction and the usage of better localized filters. On the other hand, this extra interference can be dealt with a detection structure with inherent interference cancellation, such as the considered one in the filtered time domain. This allows the full potential of this waveform, with good \ac{OOBE} characteristics, spectral efficiency and error performance at the same time.

Future work involves the expansion of the analysis presented here to message passing detection, allowing low-complexity detection with interference cancellation. Also, the good SIR characteristics of the \ac{AFBM} waveform can allow an investigation of its usage in MIMO systems, where the inherent interference is always an issue in filterbank-based systems.

\balance
% \newpage
\selectlanguage{english}
\bibliographystyle{IEEEtran}
\bibliography{references.bib}

@article{rou2025affine,
  title={Affine Frequency Division Multiplexing (AFDM) for 6G: Properties, Features, and Challenges},
  author={Rou, Hyeon Seok and Ranasinghe, Kuranage Roche Rayan and Savaux, Vincent and de Abreu, Giuseppe Thadeu Freitas and Masouros, Christos and others},
  journal={arXiv preprint arXiv:2507.21704},
  year={2025}
}

@INPROCEEDINGS{Ranasinghe_ICNC2025_oversampling,
  author={Rayan Ranasinghe, Kuranage Roche and Ge, Yao and Freitas de Abreu, Giuseppe Thadeu and Liang Guan, Yong},
  booktitle={2025 International Conference on Computing, Networking and Communications (ICNC)}, 
  title={Joint Channel Estimation and Data Detection for AFDM Receivers With Oversampling}, 
  year={2025},
  volume={},
  number={},
  pages={823-828},
  doi={10.1109/ICNC64010.2025.10994040}}

@misc{senger2025affinefilterbankmodulation,
      title={Affine Filter Bank Modulation: A New Waveform for High Mobility Communications}, 
      author={Henrique L. Senger and Gustavo P. Gonçalves and Bruno S. Chang and Hyeon Seok Rou and Kuranage Roche Rayan Ranasinghe and Giuseppe Thadeu Freitas de Abreu and Didier Le Ruyet},
      year={2025},
      eprint={2505.03589},
      archivePrefix={arXiv},
      primaryClass={eess.SP},
      url={https://arxiv.org/abs/2505.03589}, 
}

@ARTICLE{Rou_SPM_2024,
  author={Rou, Hyeon Seok and de Abreu, Giuseppe Thadeu Freitas and Choi, Junil and González G., David and Kountouris, Marios and Guan, Yong Liang and Gonsa, Osvaldo},
  journal={IEEE Signal Process. Mag.}, 
  title={From Orthogonal Time–Frequency Space to Affine Frequency-Division Multiplexing: A comparative study of next-generation waveforms for integrated sensing and communications in doubly dispersive channels}, 
  year={2024},
  volume={41},
  number={5},
  pages={71-86},
  doi={10.1109/MSP.2024.3422653}}

@ARTICLE{pereira2022generalized,
  author={Junior, Rogério Pereira and Rocha, Carlos A. F. da and Chang, Bruno S. and Le Ruyet, Didier},
  journal={IEEE Wireless Communications Letters}, 
  title={A Generalized {DFT} Precoded Filter Bank System}, 
  year={2022},
  volume={11},
  number={6},
  pages={1176-1180},
  doi={10.1109/LWC.2022.3160301}}

@ARTICLE{pereira2023two,
  author={Junior, Rogério Pereira and da Rocha, Carlos Aurélio Faria and Chang, Bruno S. and Le Ruyet, Didier},
  journal={IEEE Transactions on Wireless Communications}, 
  title={A Two-Dimensional {FFT} Precoded Filter Bank Scheme}, 
  year={2023},
  volume={22},
  number={11},
  pages={8366-8377},
  doi={10.1109/TWC.2023.3262442}}

@article{nissel2018pruned,
  title={Pruned {DFT-spread FBMC}: Low {PAPR}, low latency, high spectral efficiency},
  author={Nissel, Ronald and Rupp, Markus},
  journal={IEEE Transactions on Communications},
  volume={66},
  number={10},
  pages={4811--4825},
  year={2018},
  publisher={IEEE}
}

@ARTICLE{pereira2021novel,
  author={Junior, Rogério Pereira and Rocha, Carlos A. F. da and Chang, Bruno S. and Le Ruyet, Didier},
  journal={IEEE Wireless Communications Letters}, 
  title={A Novel {DFT} Precoded Filter Bank System With Iterative Equalization}, 
  year={2021},
  volume={10},
  number={3},
  pages={478-482},
  doi={10.1109/LWC.2020.3035332}}

@INPROCEEDINGS{shen2025timedomain,
  author={Shen, Cheng and Yuan, Jinhong},
  booktitle={ICC 2025 - IEEE International Conference on Communications}, 
  title={Time Domain Zero-Postfix (TZP) AFDM with Two-Stage Iterative MMSE Detection}, 
  year={2025},
  volume={},
  number={},
  pages={4780-4785},
  doi={10.1109/ICC52391.2025.11161127}}

@INPROCEEDINGS{wen2022downlink,
  author={Wen, Haifeng and Yuan, Weijie and Li, Shuangyang},
  booktitle={2022 IEEE International Conference on Communications Workshops (ICC Workshops)}, 
  title={Downlink OTFS Non-Orthogonal Multiple Access Receiver Design based on Cross-Domain Detection}, 
  year={2022},
  volume={},
  number={},
  pages={928-933},
  doi={10.1109/ICCWorkshops53468.2022.9814572}}

@Article{liu2025entropy,
AUTHOR = {Liu, Mengmeng and Li, Shuangyang and Bai, Baoming and Caire, Giuseppe},
TITLE = {Cross-Domain OTFS Detection via Delay–Doppler Decoupling: Reduced-Complexity Design and Performance Analysis},
JOURNAL = {Entropy},
VOLUME = {27},
YEAR = {2025},
NUMBER = {10},
ARTICLE-NUMBER = {1062},
URL = {https://www.mdpi.com/1099-4300/27/10/1062},
PubMedID = {41149020},
ISSN = {1099-4300},
DOI = {10.3390/e27101062}
}

@article{das2020time,
  title={Time domain channel estimation and equalization of CP-OTFS under multiple fractional Dopplers and residual synchronization errors},
  author={Das, Suvra Sekhar and Rangamgari, Vivek and Tiwari, Shashank and Mondal, Subhas Chandra},
  journal={IEEE Access},
  volume={9},
  pages={10561--10576},
  year={2020},
  publisher={IEEE}
}

@article{raviteja2018practical,
  title={Practical pulse-shaping waveforms for reduced-cyclic-prefix OTFS},
  author={Raviteja, Patchava and Hong, Yi and Viterbo, Emanuele and Biglieri, Ezio},
  journal={IEEE Transactions on Vehicular Technology},
  volume={68},
  number={1},
  pages={957--961},
  year={2018},
  publisher={IEEE}
}

@misc{ranasinghe2025affinefilterbankmodulation,
      title={Affine Filter Bank Modulation (AFBM): A Novel 6G ISAC Waveform with Low PAPR and OOBE}, 
      author={Kuranage Roche Rayan Ranasinghe and Henrique L. Senger and Gustavo P. Gonçalves and Hyeon Seok Rou and Bruno S. Chang and Giuseppe Thadeu Freitas de Abreu and Didier Le Ruyet},
      year={2025},
      eprint={2509.05683},
      archivePrefix={arXiv},
      primaryClass={eess.SP},
      url={https://arxiv.org/abs/2509.05683}, 
}

\end{document}